\documentclass{PoS}
\usepackage{color,graphicx}
\usepackage{amsmath}

\title{Nucleon Femtography from Exclusive Reactions}

\ShortTitle{Nucleon Femtography}

\author{\speaker{Simonetta Liuti}\thanks{This work was done under DOE Grant DE-SC0016286}\\
        University of Virginia \\
        E-mail: \email{sl4y@virginia.edu}}

\abstract{Major breakthroughs over the last two decades have led us to access information on how the nucleon's mass, spin and  mechanical properties are generated from its quark and gluon degrees of freedom. On one side, a theoretical framework has been developed which enables the extraction of 3D parton distributions from deeply virtual exclusive scattering experiments. On the other hand, the so called gravitomagnetic form factors parameterizing the QCD energy momentum tensor of the nucleon have been connected to the Mellin moments of the 3D parton distributions. Current efforts in both experiment and theory are being directed at using information from electron scattering experiments to map out and eventually visualize these 3D distributions as well as the mass, spin and  mechanical properties of the nucleon and of spin 0, 1/2, and 1 nuclei. A new science of nucleon femtography is emerging where the 3D structure of the nucleon will be studied merging information from current and forthcoming data from various facilities including the future EIC and the new influx of data science, imaging, and visualization.}
\FullConference{23rd International Spin Physics Symposium - SPIN2018 -\\
		10-14 September, 2018\\
		Ferrara, Italy}

\begin{document}
\section{Introduction}
The overarching goal of several experimental programs at Jefferson Lab as well as at the planned Electron Ion Collider (EIC) \cite{Accardi:2012qut,Geesaman:2015fha}, is to provide an avenue to quantitatively access the spatial 3D quark and gluon structure of the nucleon and to eventually map out its entire phase space/Wigner distribution. Knowledge on both the spatial and momentum distributions of quarks and gluons inside the nucleon will be conducive to understanding, within quantum chromodynamics (QCD), the mechanical properties of all strongly interacting matter including the mass, energy density, angular momentum, pressure and shear force distributions in both momentum and coordinate space.	
In particular, this will provide a path to ultimately solve the outstanding question of the decomposition of the proton spin into its quark and gluon orbital and spin components.

The key to unlocking direct experimental access to spatial distributions of partons inside the proton was provided by Ji \cite{Ji:1996ek}, where he suggested Deeply Virtual Compton Scattering (DVCS), $ e p \rightarrow e'p' \gamma$ ($\gamma$ being a real photon) as a fundamental probe of angular momentum. 
In DVCS, the high virtuality of the exchanged photon
ensures that we are resolving the partonic structure of the proton. At the same time, since it is an exclusive experiment, by measuring the four-momentum transfer between the initial and final proton, $\Delta$ ($\Delta^2=t$), similarly to elastic scattering experiments, one can obtain information on the location of the partons inside the proton, by Fourier transformation. 
%GPDs were first introduced to define the quark and gluon angular momentum in QCD in terms of observables from lepton proton scattering experiments. 
The hadronic matrix elements for DVCS  are described in terms of Generalized Parton Distributions (GPDs), see Refs.\cite{Ji:1996ek,Radyushkin:1997ki,DMul1} and reviews in \cite{Diehl:2003ny,Belitsky:2005qn,Kumericki:2016ehc}.
GPDs represent a subset of the full 3D parton distributions, the relativistic Wigner phase space distributions \cite{Belitsky:2003nz}, that can simultaneously provide information on both the longitudinal momentum fraction, $x$,  and the transverse location of partons, ${ b}$, Fourier conjugate to $\Delta$,  inside the nucleon. Similarly to the standard (forward) Parton Distribution Functions (PDFs), one can write the Operator Product Expansion (OPE) for the GPDs in terms of twist two operators whose matrix elements are taken between initial and final proton states with different momenta. Using Lorentz symmetry, parity and time reversal invariance, one can write three types of form factors for the spin-n operator associated to electromagnetic (vector) current: $A_{n,2i}(t)$, $B_{n,2i}(t)$, $C_{n}(t)$, with $i=0,(n-1)/2$. The  latter are identified with the  Mellin moments of the GPDs \cite{Ji:1996ek,Ji:1996nm}.

On the other hand, one can parameterize the nucleon matrix element of the Energy Momentum Tensor (EMT), $T^{\mu\nu}$,  stemming directly from the QCD Lagrangian, in terms of 3 conserved form factors: $A_{q,g}$, $B_{q,g}$ and $C_{q,g}$ \cite{Ji:1996ek}.  
Although the quark and gluon terms are separately renormalization scale dependent, their sum leads to conserved quantities and it is therefore apt to represent the mechanical properties of hadronic matter.
Similarly to the definition of the classical EMT, the QCD EMT time components encode the densities and flux densities of the quark and gluon fields energy and momentum, so that by integrating over the volume and summing over the quark and gluon components, one obtains the system's total energy and momentum from the time components, $T^{0i}$, while the space, $T^{ij}$, elements can be identified with the pressure ($i=j$), and the shear forces ($i\neq j$). Evaluating the EMT matrix elements one finds that $A_{q,g}(t=0)$ represents the total quark/gluon longitudinal momentum relative to the nucleon momentum. By Fourier transformation, the form factor, $A_{q,g}(t)$, therefore provides information on how the momentum is spatially distributed inside the nucleon. Similarly, $t C_{q,g}(t)$ is identified with the pressure density distribution in the nucleon \cite{Polyakov:2002yz,Polyakov:2002wz,Polyakov:2018zvc,Polyakov:2018rew,Lorce:2018egm}.
The combination $A_{q,g}(t)+B_{q,g}(t)$ gives the quark and gluon contributions to the proton angular momentum \cite{Ji:1996ek}.

In Ref.\cite{Ji:1996ek}, Ji took the fundamental step of linking the information from the GPDs OPE form factors for $n=2$, accessible in DVCS type experiments, and 
the  elements of the QCD EMT, by showing that:
\[ A_{20}^{q,g}= A_{q,g}(t), \quad\quad B_{20}^{q,g} =
 B_{q,g}(t), \quad\quad C_{20}^{q,g} = 4 \, C_{q,g}(t)  .
\]  
This important step enables investigations of the mechanical properties of the nucleon by extracting GPDs from dedicated exclusive high energy electron scattering experiments. 

%: measuring this quantity has been identifoed as a major quest for solving the proton spin crisis.

While we can benefit from a methodology fully developed through decades to extract Parton Distribution Functions (PDFs) from a variety of inclusive scattering experiments, 
the establishment of a similar quantitative program for GPDs has just begun.
Stating the problem in a nutshell, to obtain the important additional information 
on the 3D spatial structure of parton distributions encoded in GPDs requires a variety of deeply virtual coincidence experiments,  which, at variance with their inclusive counterparts, bring in additional complications. To mention a few phenomenology issues: GPDs appear in the cross section embedded in complex  amplitudes (the so called Compton Form Factors, CFFs); the cross section has a complicated dependence on the phase  measured by the azimuthal angle,  $\phi$, between the lepton and hadron planes; due to the enhanced kinematic complexity, it is harder to separate out the leading and subdominant components in the scale, $Q^2$. Pioneering experiments  conducted mostly at HERMES and Jefferson Lab have shown that GPDs extraction from data is potentially at reach \cite{Kumericki:2016ehc}. An impressive effort is underway to  address the aforementioned complications. In particular, this will require going beyond the standard computational toolbox and addressing the computational challenges that are presently limiting our access to the proton's 3D structure. The Center for Nucleon Femtography (CNF) \cite{CNF} was recently founded at Jefferson Lab to lead and coordinate efforts in this direction. 
%%%%
%The $T_{00}$ element allows us to describe the contribution of the quarks and gluons to the proton mass \cite{Ji:1994av}.  In order to connect with various quark matter models described in the literature \cite{Baym:2017whm}, we notice that our description holds in the short distance/high density regime, where the nucleon can be considered a statistically large system. 
%Through the Operator Product Expansion (OPE) in QCD one connects the EMT form factors,  $A(t), B(t), C(t)$, in Eq.(\ref{eq:emt_param}) with the matrix elements of local twist two operators \cite{Diehl:2003ny,Belitsky:2005qn}. 

%%%%%%%%
%%%%%%%%
\section{The QCD Energy Momentum Tensor and Generalized Parton Distributions}
\noindent The QCD EMT is defined as,
\begin{equation}
T^{\mu\nu}_{QCD} = \,\,  \overline{\psi} \, \gamma^{(\mu} i \overleftrightarrow{D}^{\nu )} \psi  + F^{\mu\alpha} F_\alpha^\nu +  \frac{1}{4}g^{\mu \nu}F^2 ,
\end{equation}
where $\psi$ and $F^{\mu\alpha}$ are the quark and gluon fields, respectively, while $g^{\mu\nu}$ is the spacetime metric.
The EMT matrix element between nucleon states was parameterized in Ref.\cite{Ji:1996nm} as,
\begin{eqnarray}
\label{EMT}
\langle p' \! \mid T_{q,g}^{\mu \nu} \mid \! p \rangle &=& \bar{U}_{s'}(p') \Big[A_{q,g}(t) \gamma^{(\mu} P^{\nu)} + B_{q,g}(t) \frac{P^{( \mu} i \sigma^{\nu ) \rho} \Delta_\rho}{2M}  \nonumber \\
&+& \frac{1}{4M} C_{q,g}(t)\left(\Delta^\mu \Delta^\nu - g^{\mu \nu} \Delta^2\right) + \overline{C}_{q,g}(t) g^{\mu \nu} M \Big] U_s(p),
\label{eq:emt_param}
\end{eqnarray}
where $q$ and $g$ are the quark and gluon labels; $M$ is the nucleon mass; the initial (final) nucleon spinor is $U_s(p)$ ($\bar{U}_{s'}(p')$); %$p, s$  and $p', s'$, respectively; 
$P=(p+p')/2$, and the momentum transfer is, $\Delta=p'-p$, $t=\Delta^2<0$.
Taking the light cone components in Eq.\eqref{EMT} leads to the momentum and angular momentum sum rules, respectively given by,
\begin{eqnarray}
\label{MomSR}
&& \sum_{i=q,g} A_i   =  \epsilon_q + \epsilon_g = 1 \\
\label{JiSR}
 \frac{1}{2} && \sum_{i=q,g} (A_i + B_i)  =  J_q + J_g = \frac{1}{2}.
\end{eqnarray}
where $\epsilon_{q,g}$ denote the fractions of the proton momentum carried by quarks and gluons.
These basic constructs of the theory in turn can be accessed experimentally owing to their connection, through the operator product expansion (OPE), to the Mellin moments of specific parton distributions which parameterize both the forward ($p=p^{\prime } )$ and off-forward ($p\neq p^{\prime } $) quark and gluon correlation functions. 
The completely unintegrated off forward quark-quark correlation function is defined as the matrix element between proton states with momenta and helicities $p, \Lambda$ and $p',\Lambda'$,
\begin{eqnarray}
\label{eq:unintcorr}
W_{\Lambda' \Lambda}^\Gamma(P,k,\Delta; {\cal U}) & = & \displaystyle\frac{1}{2}\int \frac{d^4 z} {(2 \pi)^4} e^{i k\cdot z}
 \left. \langle p', \Lambda' \mid 
 \bar{\psi} \left(-\frac{z}{2}\right) \Gamma {\cal U} \psi\left(\frac{z}{2}\right)   \mid p, \Lambda \rangle \right., \end{eqnarray} 
where ${\cal U}$ is the gauge link connecting the quark operators at positions $-z/2$ and $z/2$, $\Gamma$ is a Dirac structure, $\Gamma = {\bf 1}, \gamma^5, \gamma^\mu,\gamma^\mu\gamma^5,i\sigma_{\mu\nu}$, and $k$ is the quark four-momentum. 
The Generalized Parton Distributions (GPDs) are obtained by formally integrating Eq.(\ref{eq:unintcorr}) over $k^-$ and the transverse parton momentum, $k_T$, provided that the gauge link has the appropriate form (see discussion in Ref.\cite{Raja:2017xlo}). For the leading order electromagnetic interaction/vector case the operator is  $\Gamma = \gamma^+$, and one can parameterize the correlation function as,
\begin{eqnarray}
\label{GPDvec}
W^{\gamma^+}_{\Lambda' \Lambda } &=&\frac{1}{2P^+}\overline{U}(p',\Lambda')\left[\gamma^+ H + \frac{i\sigma^{+\Delta}}{2M}E\right]U(p,\Lambda)
= H(x,\xi,t) (\delta_{\Lambda,\Lambda'} +\frac{ (\Lambda\Delta^1 +i\Delta^2)}{2M}E(x,\xi,t) \delta_{-\Lambda,\Lambda'} 
\nonumber \\
\end{eqnarray}
$H(x,\xi,t)$ and $E(x,\xi,t)$ are the GPDs which depend on the longitudinal momentum transfer between the initial and final proton, represented through the skewness parameter $\xi$, and the four-momentum transfer squared, $t$; $x$ is the light cone momentum fraction carried by the parton \cite{Belitsky:2005qn,Diehl:2003ny}. In particular, $H_{q}(x,0,0) \equiv q(x), H_g(x,0,0) \equiv g(x)$, where $q(x)$ and $g(x)$ are the unpolarized quark (antiquark) and gluon distributions. From the structure of the correlation function one can immediately see that the second Mellin moment obtained by taking the derivative with respect to the longitudinal variable in Eq.\eqref{eq:unintcorr},
\begin{eqnarray}
\langle p'\mid \overline{\psi}(0) \, \gamma^{+} i {D}^{+ } \psi(0) \mid \rangle = 2P^+ \int_{1}^{1} dx x H_q(x,\xi,t)
\end{eqnarray}
is equivalent to the matrix element of the quark contribution to the EMT in Eq.\eqref{EMT}, Eq.\eqref{MomSR}, leading to the momentum sum rule, 
\begin{eqnarray}
\label{MomSR}
A_{q,g}(t) & = & \int_0^1 dx\, x H_{q,g}(x,0,t) \equiv A_{20}^{q,g}(t) \end{eqnarray}
A similar derivation allows us to write angular momentum in terms of   moments of GPDs as, 
\begin{eqnarray}
\label{JiSR}
A_{q,g}(t) + B_{q,g}(t) &  = & \int_{0}^1 dx\, x (H_{q,g}(x,0,t)+E_{q,g}(x,0,t)) \equiv A_{20}^{q,g}(t)  + B_{20}^{q,g}(t) .
\end{eqnarray}
For $t=0$, Eq.(\ref{JiSR}) leads to the Ji sum rule \cite{Ji:1996ek}.

\noindent $A_q$(0) has been known from deep inelastic scattering experiments for a long time since it represents the quark average longitudinal momentum fraction. Its $t$ dependence, as well as the form factor $B_q(t)$ have been extracted in first pioneering model dependent studies (see {\it e.g.} \cite{Mazouz:2007aa} and references therein). The gluon moments have not yet been obtained in experiment.  

\noindent While the focus of experimental and theoretical investigations has been mostly on the angular momentum sum rule, more recent developments address the EMT form factors $C_{q,g}$ \cite{Shanahan:2018pib} and $\overline{C}_{q,g}$, \cite{Hatta:2018sqd}, respectively. 
Similarly to how $A_{q,g}$ and $B_{q,g}$ are associated with the energy, momentum and angular momentum distributions inside the proton, the  
$C_{q,g}$ form factor (also know as the $D$-term \cite{Diehl:2003ny,Belitsky:2005qn}) describes the internal mechanical 
properties of pressure and shear distributions and it is obtained 
from the spatial components of the EMT \cite{Donoghue:2001qc,
Polyakov:2002yz,Polyakov:2002wz,Polyakov:2018zvc,Polyakov:2018rew},
\begin{eqnarray}
T_{ij}(\vec{r}) = \frac{1}{M} \int \frac{d^3 \Delta}{(2 \pi)^3} \, e^{i \vec{q} \cdot \vec{r}} \left( \Delta_i \Delta_j - \Delta^2 \delta_{ij} \right) C_{q,g}(t) 
\end{eqnarray}
$C_{q,g}$ is defined as the coefficient of the skewness dependent part of the second Mellin moment which reads,
\begin{eqnarray}
\label{MomSR}
 \int_0^1 dx\, x H_{q,g}(x,\xi,t) \equiv A_{20}^{q,g}(t) + (2\xi^2) \, C_2^{q,g} \end{eqnarray}
$C_{q}$  has also been recently been extracted
from DVCS experiments at Jefferson Lab, \cite{Burkert:2018bqq}. $C_g(t)$ has not been measured. While various lattice QCD evaluations of $A_q$, $B_q$, and $C_q$ have been performed by several groups \cite{Deka:2013zha,Alexandrou:2017oeh,Hagler:2007xi} their gluon counterparts have just recently been evaluated in Ref.\cite{Shanahan:2018pib}. 

\begin{figure}
\begin{center}
\includegraphics[width=8.5cm]{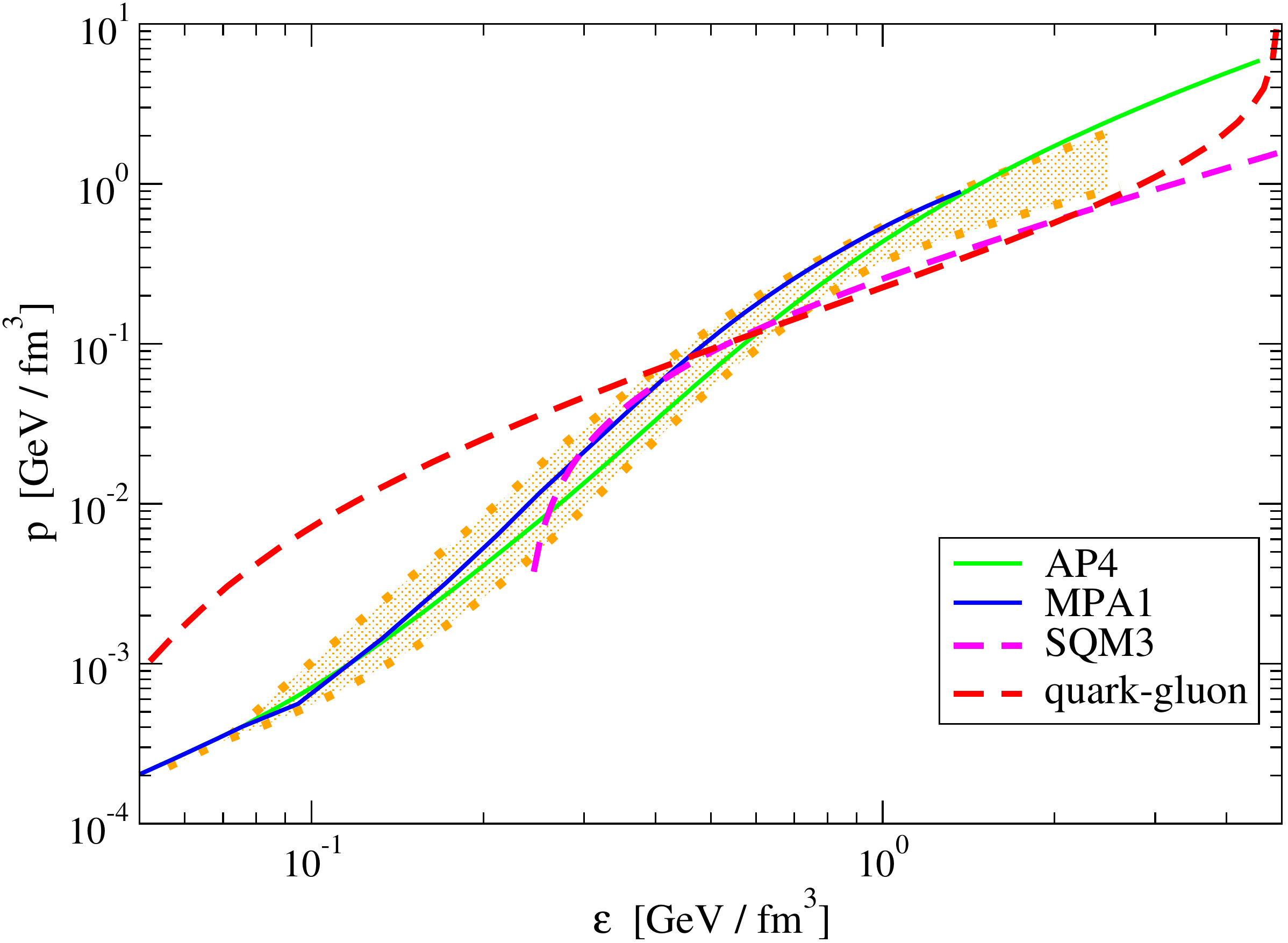}
\caption{Adapted from Ref.\cite{Liuti:2018ccr}  The EoS of neutron stars. The shaded region is the allowed region from GW170817; (red dashed) the quark-gluon EoS  constructed from the QCD EMT; two hadronic EoS AP4 (green solid) and MPA1 (blue solid); (magenta dashed) quark matter EoS SQM3. 
}
\label{fig:EoS}
\end{center}
\end{figure}

\noindent Knowing the pressure and energy-momentum spatial distributions inside the proton from the matrix elements of the QCD EMT between nucleon states has profound consequences. In particular, one can obtain constraints on the Equation of State (EoS) for the quark matter phase in hybrid neutron stars: through our QCD description of the EMT, the observables from the binary neutron stars merger  GW170817 \cite{TheLIGOScientific:2017qsa} could be confronted with laboratory measurements from electron scattering experiments \cite{Liuti:2018ccr}. In particular, the QCD-EMT-based underlying description allows us to evaluate the so far elusive gluon contribution to the EoS. In Fig.\ref{fig:EoS} the  quark-gluon contribution to the EoS constructed from the QCD EMT is shown along with different predictions and with the region allowed by the LIGO constraint from the GW170817 binary neutron star merger.
%%%%%% FIGURE EOS
%%%%%%%%%%%%%%%%%%%%%%%%%%%%%%%%%%

\noindent The  term,  $\overline{C}_{q,g}$, is
the most elusive one since it is not associated with any GPD. In Refs.\cite{Hatta:2018ina,Hatta:2018sqd} however, $\overline{C}_{g}$ was evaluated and shown to contribute to the twist-four operator $F^2$, Eq.\eqref{EMT}, or to the trace anomaly. This term can be measured in exclusive production of heavy
quarkonium states such as $J/psi$ at threshold in electron-proton scattering (\cite{Hafidi:2017bsg} and references therein).

Spin 0 nuclei are extremely interesting to study in this context because they allow us to focus on the pressure and energy terms without complications due to the nucleon spin. 
\noindent The EMT for a spin 0 system (for instance the $^4$He nucleus) reads,
\begin{eqnarray}
\label{EMT0}
\langle p' \! \mid T_{q,g}^{\mu \nu} \mid \! p \rangle &=&  \Big[A_{q,g}(t) P^\mu P^{\nu} + \frac{1}{4M} C_{q,g}(t)\left(\Delta^\mu \Delta^\nu - g^{\mu \nu} \Delta^2\right) + \overline{C}_{q,g}(t) g^{\mu \nu} M \Big] ,
\label{eq:emt_param}    
\end{eqnarray}
$C_{q,g}(t)$ was evaluated in $^4$He in Ref.\cite{Liuti:2005qj} using a realistic microscopic model of nuclear GPDs. Nuclear modifications were shown to affect this term proportionally to the nuclear separation energy $E$, namely, in a spin 0 nucleus with $A$ nucleons, $C_{q,g}(t) \propto A (1- E/M)$ (this is at variance with the steeper growth predicted by the liquid drop model based, back of the envelope calculation of \cite{Polyakov:2018zvc}).  Experimental measurements to disentangle this term are described in the ALERT proposal \cite{Armstrong:2017wfw}. 

\begin{figure}
    \centering
    \includegraphics[width=6.0cm]{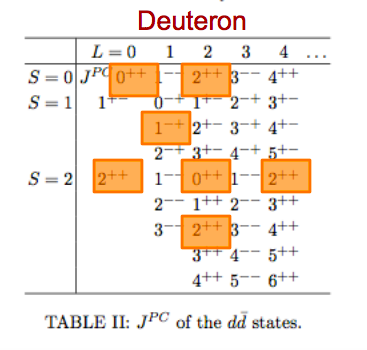}
    \caption{Adapated from \cite{SL_INT2012}. $J^{PC}$ quantum numbers of the $d\bar{d}$ states.}
    \label{fig:my_label}
\end{figure}

\noindent Finally, spin 1 systems such as the deuteron give a richer range of form factors for the EMT, essentially due to terms associated to the quadrupole moment, and to new GPDs associated to the deuteron tensor structure. The complete parameterization of the EMT tensor was described in Ref.\cite{Taneja:2011sy}, while the parameterization of the deuteron vector and axial vector correlation function in terms of GPDs was initially given in Ref.\cite{Berger:2001zb}.
It follows that, similarly to the nucleon case, by  taking the second moments with respect to $x$ of the deuteron GPDs one can find relations between the second moments of the GPDs $H_i$ and the deuteron EMT form factors 
${\cal G}_i$ ($i=1,7$)\cite{Taneja:2011sy}, 
%\begin{subequations}
\begin{eqnarray}
&& 2 \! \!\int \!dx x [ H_{1}(x,\xi,t) \!- \frac{1}{3}  H_{5}(x,\xi,t) ] \!=  {\cal G}_{1}(t) + {\xi}^2 {\cal G}_{3}(t)   
\\
&& 2 \! \int dx x H_{2}(x,\xi,t)  =  {\cal G}_{5}(t)
\label{G5}
\\ 
&& 2 \! \int dx x H_{3}(x,\xi,t)  =  {\cal G}_{2}(t) + {\xi}^2 {\cal G}_{4}(t)  \\ 
&& - 4 \! \! \int \!dx x H_{4}(x,\xi,t) = \xi {\cal G}_{6}(t) \\
&& \! \! \int dx x H_{5}(x,\xi,t) =   -\frac{t}{8M_D^2} {\cal G}_{6}(t)  + \frac{1}{2} {\cal G}_{7}(t)
\label{eq:GPDFFdeu}
\end{eqnarray}
Based on counting of the $J^{PC}$ quantum numbers in the t-channel (Figure \ref{fig:my_label} and Ref.\cite{SL_INT2012}), the EMT admits 7 conserved  deuteron form factors (see discussion in \cite{Hagler:2004yt}). The physical meaning of ${\cal G}_{5}(t)$ entering Eq.(\ref{G5}) is particularly interesting since it connects partonic angular momentum in the deuteron, $J_z$, with the GPD $H_2$, which represents the magnetic response,  
\begin{equation}
\langle p'|\int d^{3}x (\vec{x} \times \vec{T}_{q,g}^{0 i})_{z} |p\rangle
= G_{5,2}(0) \int d^{3}x ~ p^{0} \Rightarrow G_{5,2}(0) = 2 {J}^{q}_z
\end{equation}
%%%%%%%%%%%%%%%%%%%%%%%%%%%%%%%%%%%%%%%%%%%%%%%%%%%%%%%%%%%%%%%%%%%%
%%%%%%%%%%%%%%%%%%%%%%%%%%%%%%%%%%%%%%%%%%%%%%%%%%%%%%%%%%%%%%%%%%%%
%%%%%%%%%%%%%%%%%%%%%%%%%%%%%%%%%%%%%%%%%%%%%%%%%%%%%%%%%%%%%%%%%%%%

\subsection{OAM and other generalized Wandzura Wilczek  relations}
\label{sec:OAM}
 An extensive amount of literature has been dedicated to understanding how the quark and gluon angular momentum is partitioned into its spin and orbital components. 
While the momentum sum rule has an immediate dynamical interpretation in terms of the average longitudinal momentum carried by the different parton components, to obtain a dynamically transparent expression for the angular momentum sum rule one has to break it down into its spin and Orbital Angular Momentum (OAM) components,
while simultaneously preserving the gauge invariance of the theory. The decomposition can be performed within two different approaches,  by Jaffe and Manohar (JM) \cite{Jaffe:1989jz},
\begin{eqnarray}
\label{eq:JM_OAM}
&& \frac{1}{2}\Delta \Sigma_q + L_q^{JM} + \Delta G + L_g^{JM} = \frac{1}{2} 
\end{eqnarray}
and by Ji \cite{Ji:1996ek},
\begin{eqnarray}
&& \frac{1}{2}\Delta \Sigma_q + L_q^{Ji} + J_g^{Ji} = \frac{1}{2}\quad .
\label{eq:Ji_OAM}
\end{eqnarray}
The JM longitudinal OAM distribution has been identified with a parton Wigner distribution weighted by the cross product of position and momentum in the transverse plane, $b_T\times k_T$ \cite{Lorce:2011kd}. Parton Wigner distributions can be related, through Fourier transformation, to specific Generalized Transverse Momentum-Dependent Parton Distributions (GTMDs), which are off-forward TMDs. 
The correlation defining OAM corresponds to the GTMD  $F_{14}$ (we follow the naming scheme of Ref.\cite{Meissner:2009ww}). 
In particular, the OAM distribution is described by the $x$-dependent $k_T^2$ moment of $F_{14}$.  
The difference between JM and Ji's decompositions arises from the way in which the gauge invariance of the theory intervenes through the gauge link in the relevant parton correlator from which $F_{14} $ is extracted \cite{Hatta:2011ku}. Ji OAM results from a straight gauge link, whereas JM OAM results from a staple-shaped gauge link such as the one used in the standard definition of TMDs. In Ref.\cite{Burkardt:2012sd}, in particular, it was shown that JM OAM, $L_q^{JM}$, can be written as the sum of Ji's OAM, $L_q^{Ji}$, plus a matrix element including the gluon field. While $J_{q,g}$ measurements through GPDs are in progress, 
GTMDs have, however, only been evaluated in ab initio calculations \cite{Engelhardt:2017miy}, since in order to disentangle the $k_T$ and $b_T$ (or $\Delta_T$) directions require so far prohibitive exclusive measurements of two outgoing particles in the two distinct hadronic planes. 

%Regardless of the gauge link structure,  
\noindent Refs.\cite{Raja:2017xlo,Rajan:2016tlg} were dedicated to defining observables that would enable direct access to OAM in experimental measurements. In particular, the following relations which combine information from Lorentz Invariant Relations (LIRs) and QCD Equations of Motion (EoM) relations, allow us to write the GTMD defining OAM in terms of a twist-3 GPD, $\tilde{E}_{2T}$,
\begin{equation}
\label{eq:LIR3_alt}
L_z^q(x) = - \int d^2 k_T  \, \frac{ k_T^2}{M^2} \, F_{14} =
 \int_x^1 dy \, \left( \widetilde{E}_{2T} + H + E + {\cal A}_{F_{14} } \right)
\end{equation}
where we took the forward limit ($\xi\rightarrow 0$); ${\cal A}_{F_{14} } (x)$ is a term containing the gauge link dependent, or quark-gluon-quark, components of the correlation function (for a straight gauge link, ${\cal A}_{F_{14} }(x)=0$).
Other relations important for OAM are,
\begin{eqnarray}
(L_z S_z)_q(x) = \int d^2k_T \, \frac{k_T^2}{M^2} \, G_{11}  & = & \int_x^1 dy \, \left(2\widetilde{H}_{2T}' + E_{2T}'
  + \widetilde{H}- {\cal A}_{G_{11} } \right)
  \label{eq:pres_so} 
  \\
\int d^2 k_T \, \frac{k_T^2}{M^2}  \, F_{12}^{o} &  \equiv & - f_{1T}^{\perp (1)}  = -\left. {\cal M}_{F_{12}} \right|_{\Delta_{T} =0}
\label{pres_sivers}
\end{eqnarray}
Eq.\eqref{eq:pres_so} represents the spin-orbit contribution, while in Eq.\eqref{pres_sivers}, the {\em lhs} is an off-forward generalization of the Sivers shift \cite{Sivers:1989cc,Engelhardt:2015xja,Yoon:2017qzo}, while the term ${\cal M}_{F_{12}}$ on the {\it rhs} is  an off-forward/generalized analogue of the Qiu-Sterman $T_q(x,x)$ term \cite{Qiu:1991pp}.

%%%%%%%%%%%%%%%%%%%%%%%%%%%%%%%%%%%%%%%%%%%%%%%%%%%%%%%
%%%%%%%%%%%%%%%%%%%%%%%%%%%%%%%%%%%%%%%%%%%%%%%%%%%%%%%
%%%%%%%%%%%%%%%%%%%%%%%%%%%%%%%%%%%%%%%%%%%%%%%%%%%%%%%
%%%%%%%%%%%%%%%%%%%%%%%%%%%%%%%%%%%%%%%%%%%%%%%%%%%%%%%
\section{Femtography}
The exploration of the 3D structure of the nucleon including its mechanical properties encoded in the EMT has recently given origin to the science of Nuclear Femtography. Nuclear Femtography is probed in various deeply virtual exclusive reactions: an important role is played by studying the response of the system to polarization of both the electron beam and of the nucleon and nuclear targets. The study of femtography implies a profound reorganization of our approach to exploring the internal structure of the nucleon. This step up in phenomenological methodologies is required by the complexity of the processes which is translated into many more variables that need to be simultaneously accounted for in high energy coincidence reactions. An additional set of  complications arises in extracting the various observables which appear in the cross section at the amplitude level. %
%To increase the reach of this emergent field there is an urgent need to increase the depth and speed of the gravitational wave algorithms that have enabled these groundbreaking discoveries. 
%
A new generation of current and planned experiments at the future EIC \cite{Accardi:2012qut} could in principle allow us to incorporate all the information from data and phenomenology into 
a tomographic image connecting the deepest part of the quantum world with what we see as everyday matter around us. However,
to harness and organize information from experiment and  increase the reach of this emergent field will require going beyond the standard computational toolbox.
% and  addressing the computational challenges that
%are presently limiting our access to the proton's 3D structure. %which is not big enough to build a complete framework to carry out the goals.

The Center for Nuclear Femtography \cite{CNF} was  recently created at Jefferson Lab to address this diverse set of issues. On one side, there is a need for a new computational and visualization effort to 
examine and evaluate the use of new state of the art computational methods and 
techniques, including visualization to address the many layers of analysis which are 
necessary to extract the signal in its complex background after the large experimental  
data sets are acquired.
On the other, several theoretical issues are being addressed from modeling the 3D structure to providing a full description of the cross sections for the various deeply virtual exclusive processes is needed. 

An essential starting point is to provide the formalism and theoretical framework for deeply virtual exclusive-type experiments including DVCS, Deeply Virtual Meson Production, $ep \rightarrow e'p' M$ (DVMP), and Timelike Compton Scattering (TCS), $\gamma p \rightarrow {\it l}^+ {\it l}^- p'$, where a large invariant mass lepton pair is produced (recently reviewed in \cite{Kumericki:2016ehc}). Only by measuring the polarization of the beam (including the photon beam in TCS), of the target and outgoing protons will provide an adequate number of observables to extract the various Compton Form Factors containing information on GPDs. For a reliable extraction and interpretation of physics observables from experiment it is important to introduce the formalism for all deeply virtual exclusive processes according to an agreed-upon-by-the-community set of benchmarks   

A complete formalism for the photon electroproduction cross section including DVCS and the interference with the Bethe-Heitler (BH) process was recently provided in Ref.\cite{Kriesten:2019jep}. 
The new formalism is a step up from the  
formalism adopted in DVCS analyses so far which was describing the cross section through an approximate series in harmonics of the azimuthal angle, $\phi$. 
The formalism to treat coincidence reactions is based on a generalized  Rosenbluth-type description \cite{Rosenbluth:1950yq}. This formulation, while  emphasizing the physics content of the various contributions, {\it e.g.} by making a clear parallel with coincidence scattering experiments, introduces more complex $\phi$ dependent kinematic coefficients than the harmonics based  prescription.
This level of precision is a must in order to extract GPDs from DVCS analyses in order to disentangle the twist two contributions from transversity gluons and  higher twist terms.
The price of evaluating more complex $\phi$ structures is payed off not only by having a much clearer physics-based formulation, but also by the fact that the coefficients are exactly calculable. 
For illustration, we show the contributions of the  BH  and  BH-DVCS interference term  to the unpolarized cross section,
\begin{eqnarray}
\label{eq:BHintro}
\frac{d^5\sigma^{BH}_{unpol}}{d x_{Bj} d Q^2 d|t| d\phi d\phi_S } &=&
\frac{\Gamma}{t^2} \left[ A_{BH} \left(F_1^2 + \tau F_2^2 \right)+ B_{BH} \tau G_M^2(t)  \right] \\
\label{eq:BHDVIntro}
\frac{d^5\sigma^{{\cal I}}_{unpol}}{d x_{Bj} d Q^2 d|t| d\phi d\phi_S } &=&   \frac{\Gamma}{Q^2 (- t ) } \,  \Big[ 
A_{\cal I}   \left(F_1 \Re e \mathcal{H} + \tau F_2  \Re e \mathcal{E} \right)  
+ B_\mathcal{I} \,  G_M  \, \Re e (\mathcal{H+E})
+ C_\mathcal{I} \,   
G_M \, \Re e \widetilde{\mathcal{H}} \Big] . \nonumber \\
\end{eqnarray}
The detailed equations are derived and discussed in Ref.\cite{Kriesten:2019jep} (for illustration purposes, only the unpolarized case  for BH (Eq.\eqref{eq:BHintro}) and the unpolarized leading order  for the BH-DVCS interference term (labeled ${\cal I}$ in Eq.\eqref{eq:BHDVIntro}), are quoted). 
%%%%%%%%%%%%%%%%%%%%% FIGURE 2
\begin{figure}
%\hspace{2cm}
\includegraphics[width=10cm]{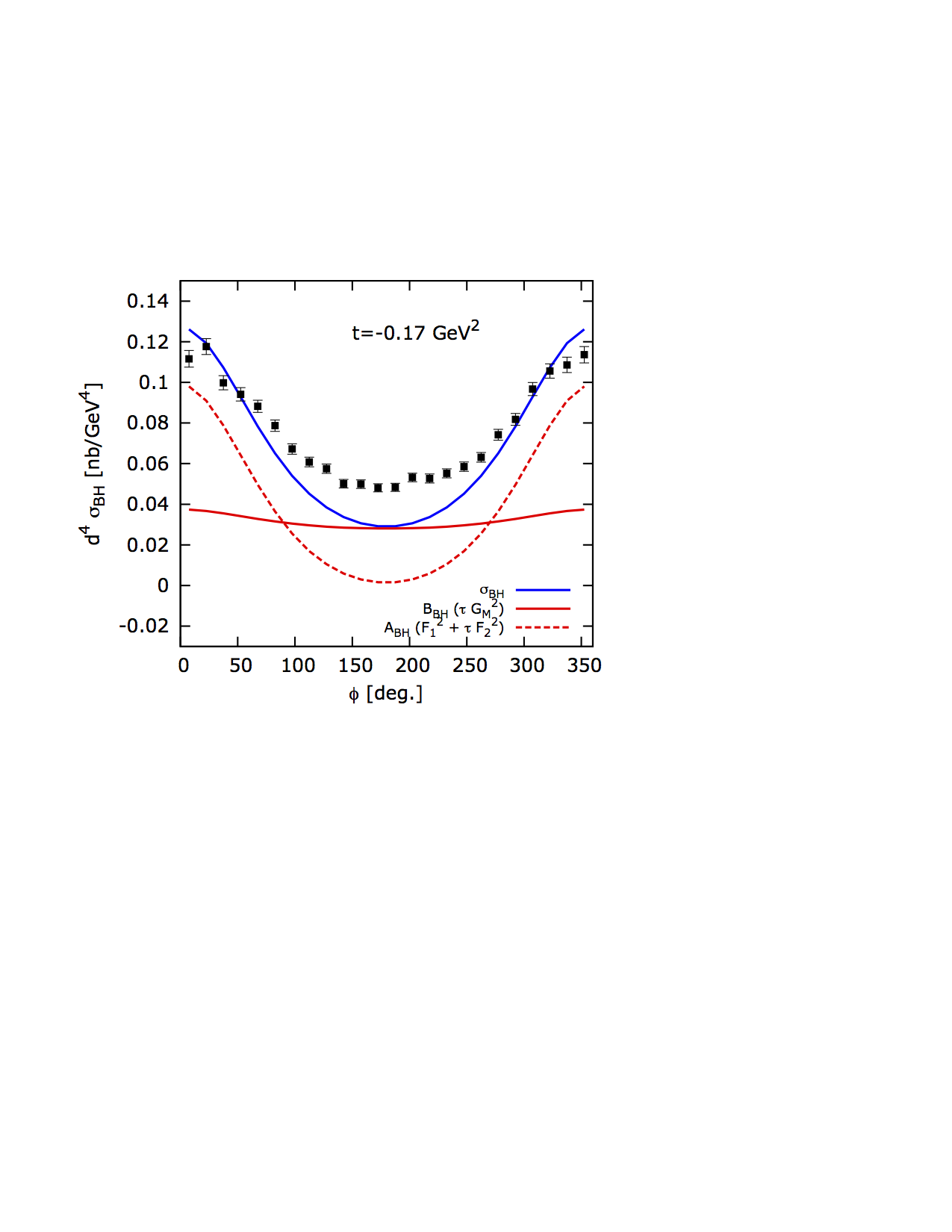}
\hspace{-3cm}
\includegraphics[width=9.5cm]{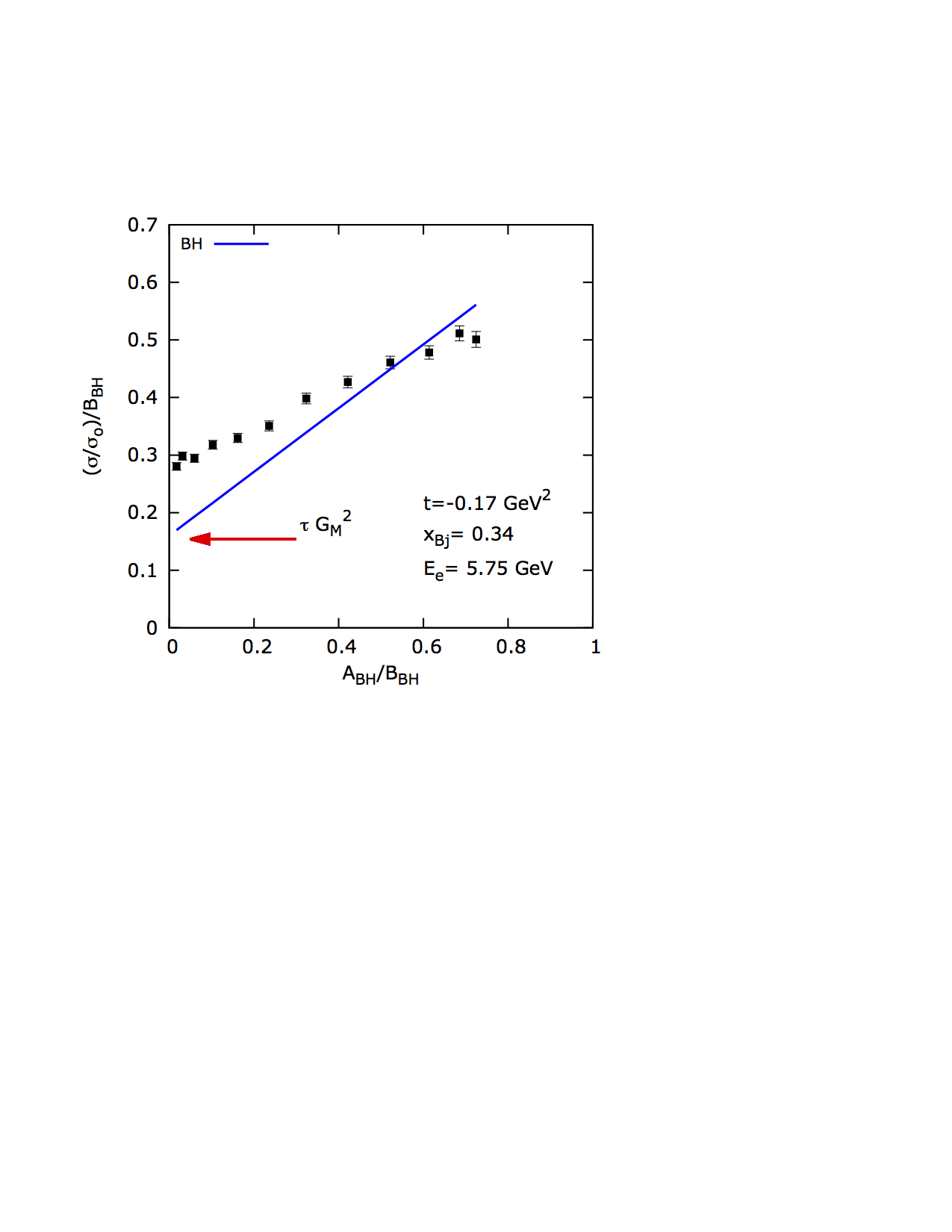}
\vspace{-5cm}
\caption{Adapted from Ref.\cite{Kriesten:2019jep}. Left: A sample of the new precise $ep\rightarrow e'p' \gamma$ data from Ref.\cite{Defurne:2015kxq} for unpolarized lepton proton scattering in the kinematic setting: $E_e=5.75$ GeV, $Q^2 = 1.82$ GeV$^2$, $x_{Bj}=0.34$, $t=-0.172$ GeV$^2$. The cross section is plotted vs. the azymuthal angle, $\phi$. The curves represent the BH contribution: (red dashed line) $\tau G_M^2$ term; (red full line) $(F_1^2 + \tau F_2^2)$ term; (blue) sum of the two; Right: Reduced cross section obtained from the same set of data  plotted vs. the kinematic variable $A_{BH}/B_{BH}$ from the Rosenbluth-type formula, Eq.(\ref{eq:BHintro}). The straight line represents the BH calculation  intercepting the y-axis at $\tau G_M^2$. 
%The difference between the data points and the curve reflects the contribution from the DVCS process.The formulation of the BH cross section is given in detail in Section
}
\label{fig:defurne1}
\end{figure}
%%%%%%%%%%%%%%%%%%%%%%%%%%%%%%%%%%

In both equations, $F_1$, $F_2$ are the Dirac and Pauli form factors , $G_M=F_1+F_2$ is the magnetic form factor ($G_E^2 = F_1^2 + \tau F_2^2$); $t$ is the momentum transfer squared,  ($\tau=-t/4M^2$); in Eq.(\ref{eq:BHDVIntro}) $\mathcal{H},  \mathcal{E}, \widetilde{\cal H}$  are  Compton form factors containing the GPDs that integrate to $F_1$, $F_2$ and $G_A$, respectively \cite{Diehl:2003ny}. 
$A_{BH}, B_{BH}$ are kinematic coefficients which are exactly calculable and rendered in covariant form; $A_\mathcal{I}, B_\mathcal{I}, C_\mathcal{I}$ are also covariant kinematic coefficients which, however, contain an extra dependence on the phase $\phi$ as we also explain in what follows. 
The new formalism allows one to emphasize the physics content of the cross section: in both Eq. \eqref{eq:BHintro} and Eq.\eqref{eq:BHDVIntro}  we can identify the first term with the electric form factor type contribution, and the second term with the magnetic form factor contribution. For the BH-DVCS interference we also have an extra function which includes the axial GPD (interestingly, a similar term would also be present in BH but parity violating). 
%%%%%%%%%%%%%%%%%%%%% FIGURE 4
\begin{figure}
%\hspace{2cm}
\includegraphics[width=6.5cm]{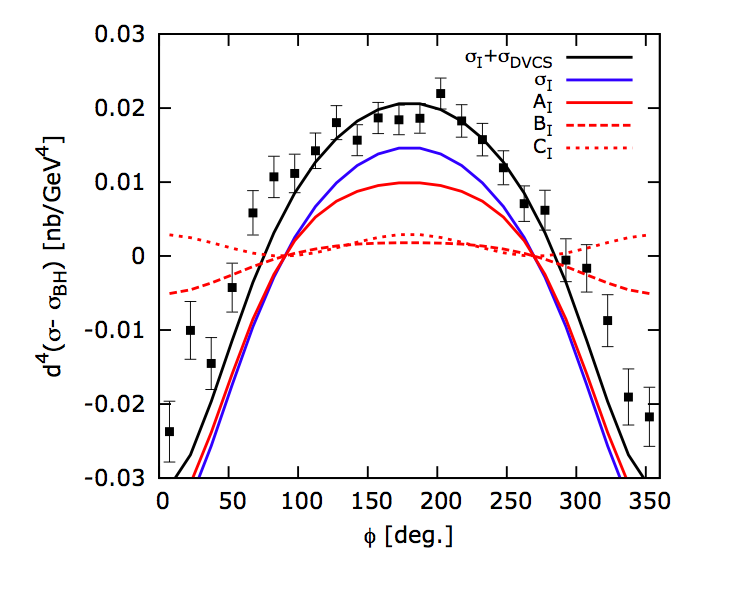}
\hspace{1cm}
\includegraphics[width=6cm]{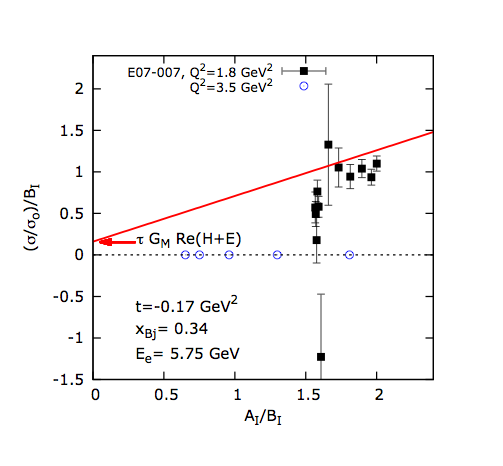}
\caption{Preliminary results on the generalized Rosenbluth separation performed for the BH-DVCS interference term displayed in Eq.\ref{eq:BHDVIntro}. Left: difference of the $ep\rightarrow e'p' \gamma$ unpolarized cross section data from Ref.\cite{Defurne:2015kxq} in the kinematic setting: $E_e=5.75$ GeV, $Q^2 = 1.82$ GeV$^2$, $x_{Bj}=0.34$, $t=-0.172$ GeV$^2$, and the calculated BH cross section; the blue line is a model calculation of the interference term; the red lines represent the separate contributions of the terms with coefficients $A_{\cal I}$, $A_{\cal I}$ and $A_{\cal I}$, respectively; the black line is the sum of the BH-DVCS interference and the DVCS modulus squared model calculations.  Right: Reduced cross section for the BH-DVCS interference term obtained from the same set of data in the Left panel plotted vs. the kinematic variable $A_{I}/B_{I}$ from the Rosenbluth-type formula, Eq.(\ref{eq:BHDVIntro}). The straight line represents the BH-DVCS interference calculation  intercepting the y-axis at $\tau G_M (\cal{H} + {\cal E})$. For this kinematics the data points are clustered away from the y-axis, making the determination of this term more uncertain. The open blue points are a kinematic evaluation of the $A_{I}/B_{I}$ values obtained by choosing a higher value of $Q^2$, $Q^2=3.5$ $^2$ for the same values of $E_e, x_{Bj} and t$. This example shows that by varying $Q^2$ one obtains a better coverage towards low values of $A_{\cal I}/B_{\cal I}$, which wold allow a precise extraction of the angular momentum term from data.
%The difference between the data points and the curve reflects the contribution from the DVCS process.The formulation of the BH cross section is given in detail in Section
}
\label{fig:defurne2}
\end{figure}
%%%%%%%%%%%%%%%%%%%%%%%%%%%%%%%%%%

In Figure \ref{fig:defurne1} we illustrate the working of the Rosenbluth separation, Eq.\eqref{eq:BHintro}, for one of the kinematic settings from the  Jefferson Lab experiment E00-110 \cite{Defurne:2015kxq}: on the left panel we show the $e p \rightarrow e'p'\gamma$ unpolarized cross section data plotted vs. $\phi$; on the right panel
we plot the reduced cross section for the same set of data vs. the kinematic variable, $A_{BH}/B_{BH}$ (the detailed definition of this quantity is given in \cite{Kriesten:2019jep}). The BH cross section appears as a linear function of the variable $A_{BH}/B_{BH}$, with intercept given by $\tau G_M^2$ and slope given by $F_1^2 + \tau F_2^2$. The difference between the data and the BH line reflects the contribution from the DVCS process.

A generalized Rosenbluth separation can be performed for the BH-DVCS interference case, Eq.\eqref{eq:BHDVIntro}, by defining an analogous kinematic variable, $A_{\cal I}/B_{\cal I}$ similar to $A_{BH}/B_{BH}$ defined for BH.
 As it can be surmised from Eq.(\ref{eq:BHDVIntro}), if one groups the small $C_{\mathcal I}$ coefficient with the $A_{\mathcal I}$ term, the intercept with the y-axis is given by $\tau  G_M \Re e ({\cal H}+ {\cal E})$.
We illustrate this situation in Figure \ref{fig:defurne2}.

Similar structures are found for other polarization configurations. In particular, the careful analysis conducted in \cite{Kriesten:2019jep} allows us to disentangle all the twist three contributions: among these in particular, is the GPD $\widetilde{E}_{2T}$ which measures OAM (see Section \ref{sec:OAM}). Therefore: {\em OAM can be extracted directly from deeply virtual exclusive electroproduction data.}

\section{Conclusions and Outlook}
In summary, the study of GPDs has opened a new frontier in QCD providing a new paradigm that will allow us to both penetrate and visualize the deep structure of visible matter and to answer so far inaccessible questions on the spatial distribution of subatomic matter. Simultaneously, by connecting the GPD Mellin moments to the QCD gravitomagnetic form factors provides experimental access to exploring the nature of mass, spin and the distribution of the strong force. This step revolutionizes our way of thinking phenomenology, putting it in a central place to address physics questions previously accessed only through ``thought of" experiments in lattice QCD. Furthermore, GPDs  represent a window into Wigner distributions. To observe, evaluate and interpret Wigner distributions  will require stepping up data analyses from the standard methods adopted so far, to developing new numerical/analytic/quantum computing methodologies. Unprecedented discoveries will be at reach by harnessing information from these new analyses within a concerted effort along the path begun with establishing of the Center for Nuclear Femtography \cite{CNF}.

\vspace{1cm}
\noindent 
I am indebted to my collaborators and students: Liliet Calero-Diaz, Michael Engelhardt, Gary Goldstein, Osvaldo Gonzalez-Hernandez, Dustin Keller, Brandon Kriesten, Andy Meyer, Abha Rajan. 

%%%%%%%%%%%%%%%%%%%%%%%%%%%%%%%%%%%%%%%%%%%%%%%%%%%%%%%%%%%%%%%%%%%%%
%%%%%%%%%%%%%%%%%%%%%%%%%%%%%%%%%%%%%%%%%%%%%%%%%%%%%%%%%%%%%%%%%%%%%
%%%%%%%%%%%%%%%%%%%%%%%%%%%%%%%%%%%%%%%%%%%%%%%%%%%%%%%%%%%%%%%%%%%%%


\begin{thebibliography}{99}
\bibitem{Accardi:2012qut} 
  A.~Accardi {\it et al.},
  %``Electron Ion Collider: The Next QCD Frontier : Understanding the glue that binds us all,''
  Eur.\ Phys.\ J.\ A {\bf 52}, no. 9, 268 (2016)
  doi:10.1140/epja/i2016-16268-9
  [arXiv:1212.1701 [nucl-ex]].
\bibitem{Geesaman:2015fha} 
  A.~Aprahamian {\it et al.},
  ``Reaching for the horizon: The 2015 long range plan for nuclear science,''
  \bibitem{Ji:1996ek} 
  X.~D.~Ji,
  %``Gauge-Invariant Decomposition of Nucleon Spin,''
  Phys.\ Rev.\ Lett.\  {\bf 78}, 610 (1997)
  doi:10.1103/PhysRevLett.78.610

\bibitem{Radyushkin:1997ki} 
  A.~V.~Radyushkin,
  %``Nonforward parton distributions,''
  Phys.\ Rev.\ D {\bf 56}, 5524 (1997)
  doi:10.1103/PhysRevD.56.5524

\bibitem{DMul1} D.~Muller, D.~Robaschik, B.~Geyer, F.~M.~Dittes and J.~Horejsi,
Fortsch.\ Phys.\  {\bf 42}, 101 (1994)

\bibitem{Diehl:2003ny} 
  M.~Diehl,
  %``Generalized parton distributions,''
  Phys.\ Rept.\  {\bf 388}, 41 (2003)
  doi:10.1016/j.physrep.2003.08.002, 10.3204/DESY-THESIS-2003-018

\bibitem{Belitsky:2005qn} 
  A.~V.~Belitsky and A.~V.~Radyushkin,
  %``Unraveling hadron structure with generalized parton distributions,''
  Phys.\ Rept.\  {\bf 418}, 1 (2005)
  doi:10.1016/j.physrep.2005.06.002

 \bibitem{Kumericki:2016ehc}
 K.~Kumericki, S.~Liuti and H.~Moutarde,
  Eur.\ Phys.\ J.\ A {\bf 52}, no. 6, 157 (2016)

\bibitem{Belitsky:2003nz} 
  A.~V.~Belitsky, X.~d.~Ji and F.~Yuan,
  %``Quark imaging in the proton via quantum phase space distributions,''
  Phys.\ Rev.\ D {\bf 69}, 074014 (2004)
  doi:10.1103/PhysRevD.69.074014

\bibitem{Ji:1996nm} 
  X.~D.~Ji,
  %``Deeply virtual Compton scattering,''
  Phys.\ Rev.\ D {\bf 55}, 7114 (1997)
  doi:10.1103/PhysRevD.55.7114

\bibitem{Polyakov:2002yz} 
  M.~V.~Polyakov,
  %``Generalized parton distributions and strong forces inside nucleons and nuclei,''
  Phys.\ Lett.\ B {\bf 555}, 57 (2003)
  doi:10.1016/S0370-2693(03)00036-4
  
\bibitem{Polyakov:2002wz} 
  M.~V.~Polyakov and A.~G.~Shuvaev,
  %``On'dual' parametrizations of generalized parton distributions,''
  hep-ph/0207153.

\bibitem{Polyakov:2018zvc} 
  M.~V.~Polyakov and P.~Schweitzer,
  %``Forces inside hadrons: pressure, surface tension, mechanical radius, and all that,''
  Int.\ J.\ Mod.\ Phys.\ A {\bf 33}, no. 26, 1830025 (2018)
  doi:10.1142/S0217751X18300259

\bibitem{Polyakov:2018rew} 
  M.~V.~Polyakov and P.~Schweitzer,
  %``Mechanical properties of particles,''
  arXiv:1812.06143 [hep-ph].

\bibitem{Lorce:2018egm} 
  C.~Lorc\'{e}, H.~Moutarde and A.~P.~Trawinski,
  %``Revisiting the mechanical properties of the nucleon,''
  Eur.\ Phys.\ J.\ C {\bf 79}, no. 1, 89 (2019)
  doi:10.1140/epjc/s10052-019-6572-3

\bibitem{CNF} https://pages.shanti.virginia.edu/Femtography/

\bibitem{Mazouz:2007aa} 
  M.~Mazouz {\it et al.} [Jefferson Lab Hall A Collaboration],
  %``Deeply virtual compton scattering off the neutron,''
  Phys.\ Rev.\ Lett.\  {\bf 99}, 242501 (2007)
  doi:10.1103/PhysRevLett.99.242501


\bibitem{Donoghue:2001qc} 
  J.~F.~Donoghue, B.~R.~Holstein, B.~Garbrecht and T.~Konstandin,
  %``Quantum corrections to the Reissner-Nordstrom and Kerr-Newman metrics,''
  Phys.\ Lett.\ B {\bf 529}, 132 (2002)
  Erratum: [Phys.\ Lett.\ B {\bf 612}, 311 (2005)]

\bibitem{Burkert:2018bqq} 
  V.~D.~Burkert, L.~Elouadrhiri and F.~X.~Girod,
  %``The pressure distribution inside the proton,''
  Nature {\bf 557}, no. 7705, 396 (2018).
 
\bibitem{Deka:2013zha} M.~Deka {\it et al.},
  %``Lattice study of quark and glue momenta and angular momenta in the nucleon,''
  Phys.\ Rev.\ D {\bf 91}, no. 1, 014505 (2015)

\bibitem{Alexandrou:2017oeh} 
  C.~Alexandrou, M.~Constantinou, K.~Hadjiyiannakou, K.~Jansen, C.~Kallidonis, G.~Koutsou, A.~Vaquero Avilés-Casco and C.~Wiese,
  %``Nucleon Spin and Momentum Decomposition Using Lattice QCD Simulations,''
  Phys.\ Rev.\ Lett.\  {\bf 119}, no. 14, 142002 (2017)

\bibitem{Hagler:2007xi} 
  P.~Hagler {\it et al.} [LHPC Collaboration],
  %``Nucleon Generalized Parton Distributions from Full Lattice QCD,''
  Phys.\ Rev.\ D {\bf 77}, 094502 (2008)

\bibitem{Shanahan:2018pib} 
  P.~E.~Shanahan and W.~Detmold,
  %``Gluon gravitational form factors of the nucleon and the pion from lattice QCD,''
  Phys.\ Rev.\ D {\bf 99}, no. 1, 014511 (2019)
  doi:10.1103/PhysRevD.99.014511
 

\bibitem{TheLIGOScientific:2017qsa} 
  B.~P.~Abbott {\it et al.} [LIGO Scientific and Virgo Collaborations],
  %``GW170817: Observation of Gravitational Waves from a Binary Neutron Star Inspiral,''
  Phys.\ Rev.\ Lett.\  {\bf 119}, no. 16, 161101 (2017)
  doi:10.1103/PhysRevLett.119.161101

\bibitem{Liuti:2018ccr} 
  S.~Liuti, A.~Rajan and K.~Yagi,
  %``Bounds on the Equation of State of Neutron Stars from High Energy Deeply Virtual Exclusive Experiments,''
  arXiv:1812.01479 [hep-ph].

\bibitem{Hatta:2018ina} 
  Y.~Hatta and D.~L.~Yang,
  %``Holographic $J/\psi$ production near threshold and the proton mass problem,''
  Phys.\ Rev.\ D {\bf 98}, no. 7, 074003 (2018)

\bibitem{Hatta:2018sqd} 
  Y.~Hatta, A.~Rajan and K.~Tanaka,
  %``Quark and gluon contributions to the QCD trace anomaly,''
  JHEP {\bf 1812}, 008 (2018)
  doi:10.1007/JHEP12(2018)008

\bibitem{Hafidi:2017bsg} 
  K.~Hafidi, S.~Joosten, Z.~E.~Meziani and J.~W.~Qiu,
  %``Production of Charmonium at Threshold in Hall A and C at Jefferson Lab,''
  Few Body Syst.\  {\bf 58}, no. 4, 141 (2017).
  doi:10.1007/s00601-017-1305-3

\bibitem{Liuti:2005qj} 
  S.~Liuti and S.~K.~Taneja,
  %``Nuclear medium modifications of hadrons from generalized parton distributions,''
  Phys.\ Rev.\ C {\bf 72}, 034902 (2005)
  doi:10.1103/PhysRevC.72.034902
  
\bibitem{Armstrong:2017wfw} 
  W.~Armstrong {\it et al.},
  %``Partonic Structure of Light Nuclei,''
  arXiv:1708.00888 [nucl-ex].


\bibitem{Taneja:2011sy} 
  S.~K.~Taneja, K.~Kathuria, S.~Liuti and G.~R.~Goldstein,
  %``Angular momentum sum rule for spin one hadronic systems,''
  Phys.\ Rev.\ D {\bf 86}, 036008 (2012)
  doi:10.1103/PhysRevD.86.036008

\bibitem{Berger:2001zb} 
  E.~R.~Berger, F.~Cano, M.~Diehl and B.~Pire,
  %``Generalized parton distributions in the deuteron,''
  Phys.\ Rev.\ Lett.\  {\bf 87}, 142302 (2001)
  doi:10.1103/PhysRevLett.87.142302

\bibitem{SL_INT2012} S. Liuti, Talk at INT Workshop {\em``Orbital Angular Momentum in QCD"}, February 6-17, 2012,  
http://www.int.washington.edu/talks/WorkShops/int$\_$12$\_$49W/

\bibitem{Hagler:2004yt} 
  P.~Hagler,
  %``Form-factor decomposition of generalized parton distributions at leading twist,''
  Phys.\ Lett.\ B {\bf 594}, 164 (2004)
  doi:10.1016/j.physletb.2004.05.014
  
\bibitem{Jaffe:1989jz} 
  R.~L.~Jaffe and A.~Manohar,
  %``The G(1) Problem: Fact and Fantasy on the Spin of the Proton,''
  Nucl.\ Phys.\ B {\bf 337}, 509 (1990).
  doi:10.1016/0550-3213(90)90506-9

\bibitem{Lorce:2011kd}C.~Lorce and B.~Pasquini,
  %``Quark Wigner Distributions and Orbital Angular Momentum,''
  Phys.\ Rev.\ D {\bf 84}, 014015 (2011)
  doi:10.1103/PhysRevD.84.014015
 
\bibitem{Meissner:2009ww} 
  S.~Meissner, A.~Metz and M.~Schlegel,
  %``Generalized parton correlation functions for a spin-1/2 hadron,''
  JHEP {\bf 0908}, 056 (2009)
  doi:10.1088/1126-6708/2009/08/056

 \bibitem{Hatta:2011ku} 
  Y.~Hatta,
  %``Notes on the orbital angular momentum of quarks in the nucleon,''
  Phys.\ Lett.\ B {\bf 708}, 186 (2012)
  doi:10.1016/j.physletb.2012.01.024

\bibitem{Burkardt:2012sd} 
  M.~Burkardt,
  %``Parton Orbital Angular Momentum and Final State Interactions,''
  Phys.\ Rev.\ D {\bf 88}, no. 1, 014014 (2013)
  doi:10.1103/PhysRevD.88.014014

\bibitem{Engelhardt:2017miy} 
  M.~Engelhardt,
  %``Quark orbital dynamics in the proton from Lattice QCD -- from Ji to Jaffe-Manohar orbital angular momentum,''
  Phys.\ Rev.\ D {\bf 95}, no. 9, 094505 (2017)
  doi:10.1103/PhysRevD.95.094505

\bibitem{Raja:2017xlo} 
  A.~Rajan, M.~Engelhardt and S.~Liuti,
  %``Lorentz Invariance and QCD Equation of Motion Relations for Generalized Parton Distributions and the Dynamical Origin of Proton Orbital Angular Momentum,''
  Phys.\ Rev.\ D {\bf 98}, no. 7, 074022 (2018)
  doi:10.1103/PhysRevD.98.074022
 
\bibitem{Rajan:2016tlg} 
  A.~Rajan, A.~Courtoy, M.~Engelhardt and S.~Liuti,
  %``Parton Transverse Momentum and Orbital Angular Momentum Distributions,''
  Phys.\ Rev.\ D {\bf 94}, no. 3, 034041 (2016)
  doi:10.1103/PhysRevD.94.034041

\bibitem{Sivers:1989cc} 
  D.~W.~Sivers,
  %``Single Spin Production Asymmetries from the Hard Scattering of Point-Like Constituents,''
  Phys.\ Rev.\ D {\bf 41}, 83 (1990).

\bibitem{Engelhardt:2015xja} 
  M.~Engelhardt, P.~Hägler, B.~Musch, J.~Negele and A.~Schäfer,
  %``Lattice QCD study of the Boer-Mulders effect in a pion,''
  Phys.\ Rev.\ D {\bf 93}, no. 5, 054501 (2016)

\bibitem{Yoon:2017qzo} 
  B.~Yoon {\it et al.},
  %``Nucleon Transverse Momentum-dependent Parton Distributions in Lattice QCD: Renormalization Patterns and Discretization Effects,''
  Phys.\ Rev.\ D {\bf 96}, no. 9, 094508 (2017)

\bibitem{Qiu:1991pp} 
  J.~w.~Qiu and G.~F.~Sterman,
  %``Single transverse spin asymmetries,''
  Phys.\ Rev.\ Lett.\  {\bf 67}, 2264 (1991)
  
\bibitem{Kriesten:2019jep} 
  B.~Kriesten, S.~Liuti, L.~Calero-Diaz, D.~Keller, A.~Meyer, G.~R.~Goldstein and J.~O.~Gonzalez-Hernandez,
  %``Extraction of Generalized Parton Distribution Observables from Deeply Virtual Electron Proton Scattering Experiments,''
  arXiv:1903.05742 [hep-ph].

\bibitem{Rosenbluth:1950yq} 
  M.~N.~Rosenbluth,
  %``High Energy Elastic Scattering of Electrons on Protons,''
  Phys.\ Rev.\  {\bf 79}, 615 (1950).
 
\bibitem{Defurne:2015kxq} 
  M.~Defurne {\it et al.} [Jefferson Lab Hall A Collaboration],
  %``E00-110 experiment at Jefferson Lab Hall A: Deeply virtual Compton scattering off the proton at 6 GeV,''
  Phys.\ Rev.\ C {\bf 92}, no. 5, 055202 (2015)
  doi:10.1103/PhysRevC.92.055202

\end{thebibliography}
\end{document}